\documentclass[aps,prl,twocolumn,superscriptaddress]{revtex4-1}

\usepackage{graphicx}
\usepackage{epsfig}
\usepackage{epstopdf}
\usepackage{subfigure}
\usepackage{tikz}
\usetikzlibrary{calc}

\usepackage{dcolumn}
\usepackage{latexsym}
\usepackage{amssymb}
\usepackage{amsmath}
\usepackage{amsfonts}
\usepackage{wasysym}
\usepackage{bm}

\usepackage[colorlinks,bookmarks=false,citecolor=blue,linkcolor=red,urlcolor=blue]{hyperref}
\usepackage{verbatim}

\usepackage{booktabs}
\newcommand{\ra}[1]{\renewcommand{\arraystretch}{#1}}

\def\be{\begin{equation}}
\def\ee{\end{equation}}
\def\bea{\begin{eqnarray}}
\def\eea{\end{eqnarray}}

\def\vec{\mathbf}
\def\mc{\mathcal}

\usepackage{color}

\definecolor{darkblue}{rgb}{0,0.02,0.45}
\definecolor{darkred}{rgb}{0.45,0.02,0} 
\definecolor{darkgreen}{rgb}{0.02,0.45,0.0}

\def\Sq{\small{\Square}}
\def\Oct{\small{\octagon}}

\usepackage{times}

\begin{document}
\title{Resonating valence bond physics is not always governed by the shortest tunneling loops}

\author{Arnaud Ralko}\email{arnaud.ralko@neel.cnrs.fr}
\affiliation{Institut N\'eel, UPR2940, Universit\'e Grenoble Alpes et CNRS, Grenoble, FR-38042 France}
\author{Ioannis Rousochatzakis}\email{irousoch@umn.edu}
\affiliation{School of Physics and Astronomy, University of Minnesota, Minneapolis, MN 55116, USA}
\affiliation{Max Planck Institut f\"ur Physik Komplexer Systeme, N\"othnitzer Str. 38, 01187 Dresden, Germany}
\date{\today}

\begin{abstract}
It is well known that the low-energy sector of quantum spin liquids and other magnetically disordered systems is governed by short-ranged resonating valence bonds (RVB). Here we show that the standard minimal truncation to the nearest-neighbor valence bond (NNVB) basis fails completely even for systems where it should work the most, according to received wisdom. This paradigm shift is demonstrated for the quantum spin-$\frac{1}{2}$ square-kagome, where strong geometric frustration, similar to the kagome, prevents magnetic ordering down to zero temperature. The shortest tunneling events bear the strongest longer-range singlet fluctuations, leading to amplitudes that {\it do not} drop exponentially with the length of the loop $L$, and to an unexpected loop-six valence bond crystal (VBC), which is otherwise very high in energy at the minimal truncation level. The low-energy effective description gives in addition a clear example of correlated loop processes that depend not only on the type of the loop but also on its lattice embedding, a direct manifestation of the long-range nature of the virtual singlets.
\end{abstract}

\pacs{75.10.Jm,05.30.-d,05.50.+q}

\maketitle
{\it Introduction} -- 
The search for quantum spin liquids has been an active topic in condensed matter physics for many years~\cite{Anderson73,FazekasAnderson74, Anderson87, LiangDoucotAnderson88, Sandvik05, HFMBook,Diep, Sachdev92, Balents,Ramirez94}. Recently, this search has gained new impetus from the discovery~\cite{Shores05, Hiroi2001, Fak2012, Okamoto2009, Aidoudi2011,Mendels2007, deVries2009, Han12, Clark2013} of a series of layered spin-$\frac{1}{2}$ kagome antiferromagnets (AFMs), whose ideal nearest-neighbor (NN) Heisenberg limit stands out as the prime candidate for a $Z_2$  topological spin liquid~\cite{YanHuseWhite, Shollwock}. 

One of the earliest fundamental insights from the theory side is that the low-energy sector of all magnetically disordered AFMs should be governed by resonances between sufficiently short-ranged singlet or valence bond (VB) pairings~\cite{Anderson73,FazekasAnderson74, Anderson87, LiangDoucotAnderson88, Sandvik05}. Casting this very physical picture into an effective Hamiltonian has been challenging for many years~\cite{Sutherland88, RokhsarKivelson, ZengElser95, MambriniMila2000, Misguich02,*  Misguich03, Ralko2009, Ralko2010, Schwandt2010, Poilblanc2010, IoannisZ2}.  While one aspect of the problem, the non-orthogonality of the  VBs~\cite{RokhsarKivelson}, was essentially resolved recently~\cite{Ralko2009,Schwandt2010,Poilblanc2010,IoannisZ2}, the really serious problem turns out to be the truncation to the NNVB basis, a problem that goes back to Zeng and Elser~\cite{ZengElser95}. 

This is the second study that focuses on the impact of virtual singlets outside the NNVB basis to the tunneling physics. Ref.~[\onlinecite{IoannisZ2}] has dealt with the 2D kagome, where the delicate competition between different resonances leads to a VBC~\cite{MarstonZeng91, NikolicSenthil03, SinghHuse07, Vidal2010} which is very fragile~\cite{Poilblanc2010,Schwandt2010, AndreasQDM}, and virtual singlets are essential to stabilize the $Z_2$ spin liquid~\cite{IoannisZ2}.

\begin{figure*}[!t]
\includegraphics[width=0.90\textwidth,clip]{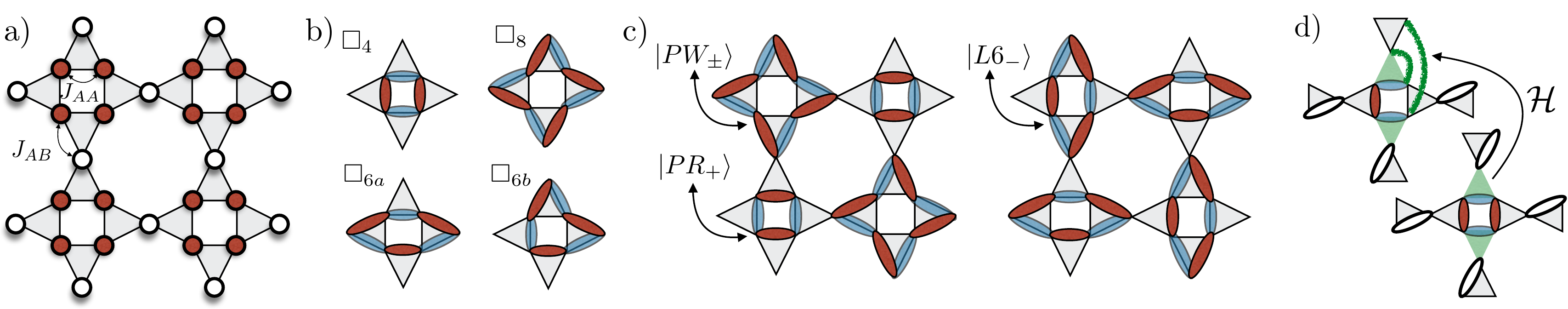}
\caption{(Color online) (a) The square-kagome  has two inequivalent sites (A and B) and two NN exchange couplings $J_{\text{AA}}$ and $J_{\text{AB}}$. (b) The dominant processes around AA-squares. Blue and red ovals denote the two NNVB states involved in each process. (c) The two types of competing VBCs, the `pinwheel' (left) and the `loop-six' (right). (d) Virtual excursions outside the NNVB basis. Here, the virtual singlet at the top is excited by applying the Hamiltonian on a defect triangle (shaded green) of one of the two NNVB states (red).}\label{fig:1}
\end{figure*}

The aim of this work is to show that virtual singlets have a strong qualitative effect in all disordered AFMs, even when the minimal NNVB truncation is expected to work the most, according to received wisdom. To this end, we have chosen an extreme case of an AFM, the square-kagome~\cite{Georges2001,Richter2006, Richter2009, IoannisKagomeLike, Hiroki2013, Derzhko2014} of Fig.~\ref{fig:1}, which features a huge tunneling energy separation at the minimal truncation level. This happens because this system manifests the shortest resonances possible, the squares [$\Sq_4$ in Fig.~\ref{fig:1}\,(b)], with NNVB amplitudes that are {\it five} times stronger than the second-shortest,  `loop-six' events. This energy separation leads to the very robust `pinwheel' VBC of Fig.~\ref{fig:1}\,(c, left)~\cite{Georges2001,IoannisKagomeLike}, well separated in energy from competing RVB states.

Yet, even for such a system, virtual singlets change the physics qualitatively because they have very different impact on processes with different loop length $L$: The square loops hold the largest density of defect triangles (triangles without VBs~\cite{Elser1989}) and thus bear much stronger quantum fluctuations, which suppress the loop-four amplitude by a striking factor of $\sim$10. This leads to `renormalized' amplitudes that {\it do not} drop exponentially with $L$, and to an unexpected `loop-six' VBC [Fig.~\ref{fig:1}\,(c, right)] which is otherwise very far in the parameter space in the absence of virtual singlets. The result that virtual singlets change the physics qualitatively even in such a system simply tells us that the minimal NNVB truncation is probably inadequate for all standard~\footnote{Certainly, one can imagine artificial cases (e.g., systems where all loops beyond the shortest are made long enough) where the minimal truncation will at some point become adequate again.} disordered AFMs.

The `dressed' NNVB description of the square-kagome offers in addition a clear example of correlated loop processes, which depend not only on the type of the loop but also on the particular VB environment of the loop. This embedding dependence is a direct manifestation of the long-range nature of virtual singlets, and adds another twist in the way we think about the effective RVB dynamics of disordered magnets.

{\it Model} -- The square-kagome features two symmetry inequivalent sites (A and B), two NN bonds (AA and AB), and two void loops (AA-squares and AB-octagons), see Fig.~\ref{fig:1}(a). Hence the most natural spin-$\frac{1}{2}$ Heisenberg model comprises two NN exchange constants, $J_{\text{AA}}$ and $J_{\text{AB}}$,
 \be
\label{eqn:Heis}
\mc{H}_{\text{Heis}}=\sum_{\langle ij\rangle} J_{ij}\vec{S}_i\cdot\vec{S}_j~,
\ee
where $\langle ij\rangle$ denotes NN sites. We target the nature of the disordered phase labeled `Y$_a$' in \cite{IoannisKagomeLike}, which appears in the highly frustrated region $x\!\equiv\!\frac{J_{\text{AB}}}{J_{\text{AA}}}\!\sim\!1$. 
This is precisely where the RVB picture is most relevant, because at $x\!=\!1$ we deal with a system of corner-sharing equilateral triangles, and such systems hold the two key RVB ingredients~\cite{ Elser1989, ZengElser95, MambriniMila2000, Misguich02,*Misguich03}: i) a manifold of energetically favorable short-ranged VB states, and ii) a  mechanism that drives the resonances between them. The first derives from the fact that each triangle can be minimized exactly by pairing two out of the three spins into a singlet. This leads to an extensive number of NNVB states where a large portion of the triangles  ($\frac{3}{4}$) host a singlet and are thus  satisfied locally. The second ingredient stems from the quantum fluctuations around the remaining, void or `defect' triangles~\cite{Elser1989}.

The square-kagome has  
22 topologically distinct dominant processes, 4 around AA-squares and 18 around AB-octagons~\cite{IoannisKagomeLike}. The former are the most crucial and are shown in Fig.~\ref{fig:1}\,(b): The loop-four or `perfect square' resonance $\Sq_4$, two loop-six processes $\Sq_{6a}$ and $\Sq_{6b}$, and finally the loop-eight or `square pinwheel' process $\Sq_8$, involving NNVB states with 2, 1 and 0 defect triangles,  respectively. Similarly, the octagonal processes~\cite{SM} consist of: the `perfect octagon' resonance $\Oct_8$, four loop-10 $\Oct_{10a-d}$, eight loop-12 $\Oct_{12a-h}$, four loop-14 $\Oct_{14a-d}$, and finally the `octagonal pinwheel' process $\Oct_{16}$, with 4, 3, 2, 1 and 0 defect triangles, respectively.

To accommodate the effect of virtual singlets on the quantum dimer model (QDM)~\cite{Sutherland88, RokhsarKivelson, ZengElser95, MambriniMila2000, Misguich02,*Misguich03, Ralko2009, Ralko2010, Schwandt2010, Trousselet2012,Hao14} we write down an {\it embedded} QDM~\cite{IoannisZ2} 
\be
\label{eq:QDM}
\mc{H}_{\text{QDM}} \!\!=\!\! \sum_{\ell, e}\! t_\ell^e \!\left(|1_\ell^e\rangle\langle2_\ell^e|\!+\!\text{h.c.}\right) 
\!+\!V_\ell^{e}\! \left(|1_\ell^e\rangle\langle1_\ell^e|\!+\!|2_\ell^e\rangle\langle2_\ell^e|\right)\!,
\ee
where we sum over all loops $\ell$ that involve a resonance between two NNVB states $|1_\ell^e\rangle$ and $|2_\ell^e\rangle$ [blue and red ovals in Fig.~\ref{fig:1}\,(b)], and all possible VB environments $e$ for given $\ell$. The corresponding tunneling amplitudes and potential energies are denoted by $t_\ell^e$ and $V_\ell^e$ (measured from $E_0\!=\!-\frac{3}{8}N$, $N$ is the number of sites).

\begin{figure*}[!tb]
\includegraphics[width=0.90\textwidth,clip]{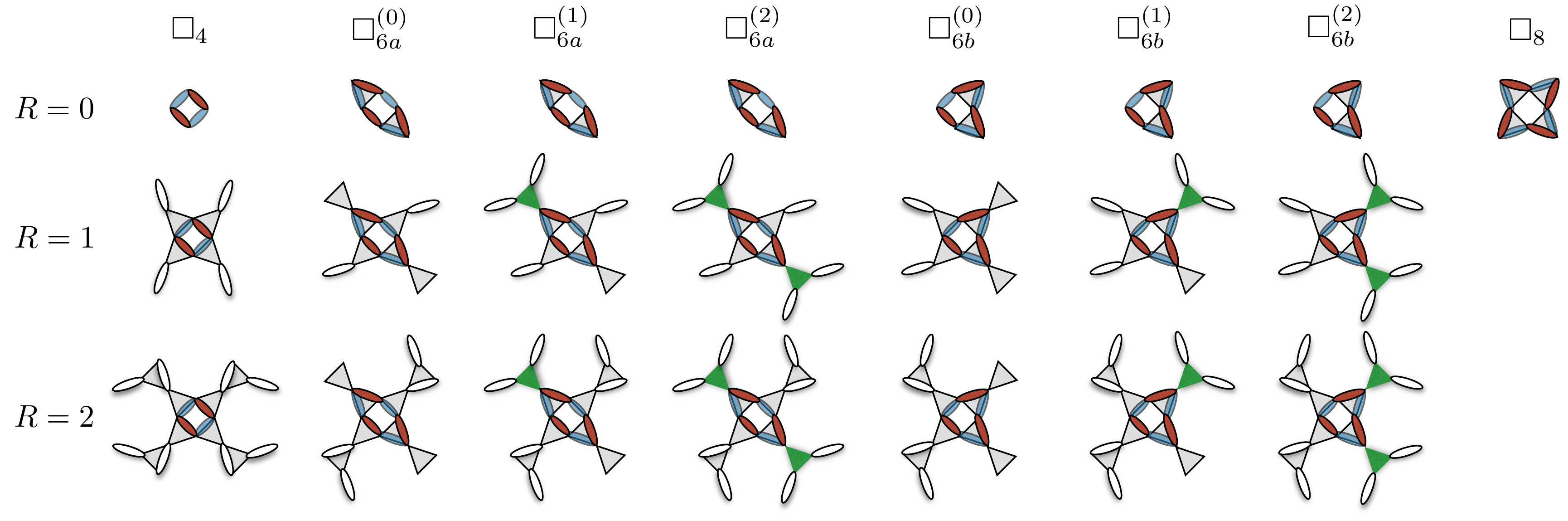}
\caption{(Color online) Finite spin-1/2 Heisenberg clusters from which we extract the tunneling amplitudes for the four dominant processes on the lattice, see text. The shaded (green) triangles indicate the closest extra defect triangles that give rise to an embedding dependence.}
\label{fig:2} 
\end{figure*}

{\it Minimal truncation \& main competing phases} -- 
Within the minimal NNVB truncation method, the parameters $t_\ell^e$ and
$V_\ell^e$ are independent of the embedding $e$~\cite{IoannisZ2}. Their values
are reported in \cite{IoannisKagomeLike} based on a method which is equivalent
to the infinite order overlap expansion of \cite{Ralko2009,Schwandt2010}.
Namely, for each process we consider a minimal cluster that can host only the
two NNVB coverings, $|1_\ell^e\rangle$ and $|2_\ell^e\rangle$, involved in the
process, see top line of Fig.~\ref{fig:2}. For these minimal clusters, the NNVB
problem amounts to diagonalizing the Hamiltonian
$\mc{H}_{\text{NNVB}}=\left(\mc{O}^{-\frac{1}{2}}\mc{H}_{\text{Heis}}
\mc{O}^{\frac{1}{2}}\right)_{\text{NNVB}}$, where $\mc{O}$ is the overlap
matrix in the NNVB basis $\{|1_\ell^e\rangle, |2_\ell^e\rangle\}$. The matrix
elements of $\mc{O}$ and $\mc{H}_{\text{NNVB}}$ can be found following standard
rules~\cite{Sutherland88,RokhsarKivelson, ZengElser95, MambriniMila2000,
Misguich02,*Misguich03,Ralko2009}. The parameters involving the AA-squares are:
$t_{\Sq_4}\!=\!-1$, $V_{\Sq_4}\!=\!\frac{1}{2}$,
$t_{\Sq_{6a}}\!=\!t_{\Sq_{6b}}\!=\!\frac{1}{5}(2x-1)$,
$V_{\Sq_{6a}}\!=\!V_{\Sq_{6b}}\!=\!\frac{1}{20}(2x-1)$, and
$t_{\Sq_8}\!=\!\frac{8}{21}(1-x)$, $V_{\Sq_8}\!=\!-\frac{1}{21}(1-x)$. For the
remaining ones see Table I of \cite{IoannisKagomeLike}.

\begin{figure}[!b]
\includegraphics[width=0.45\textwidth,clip]{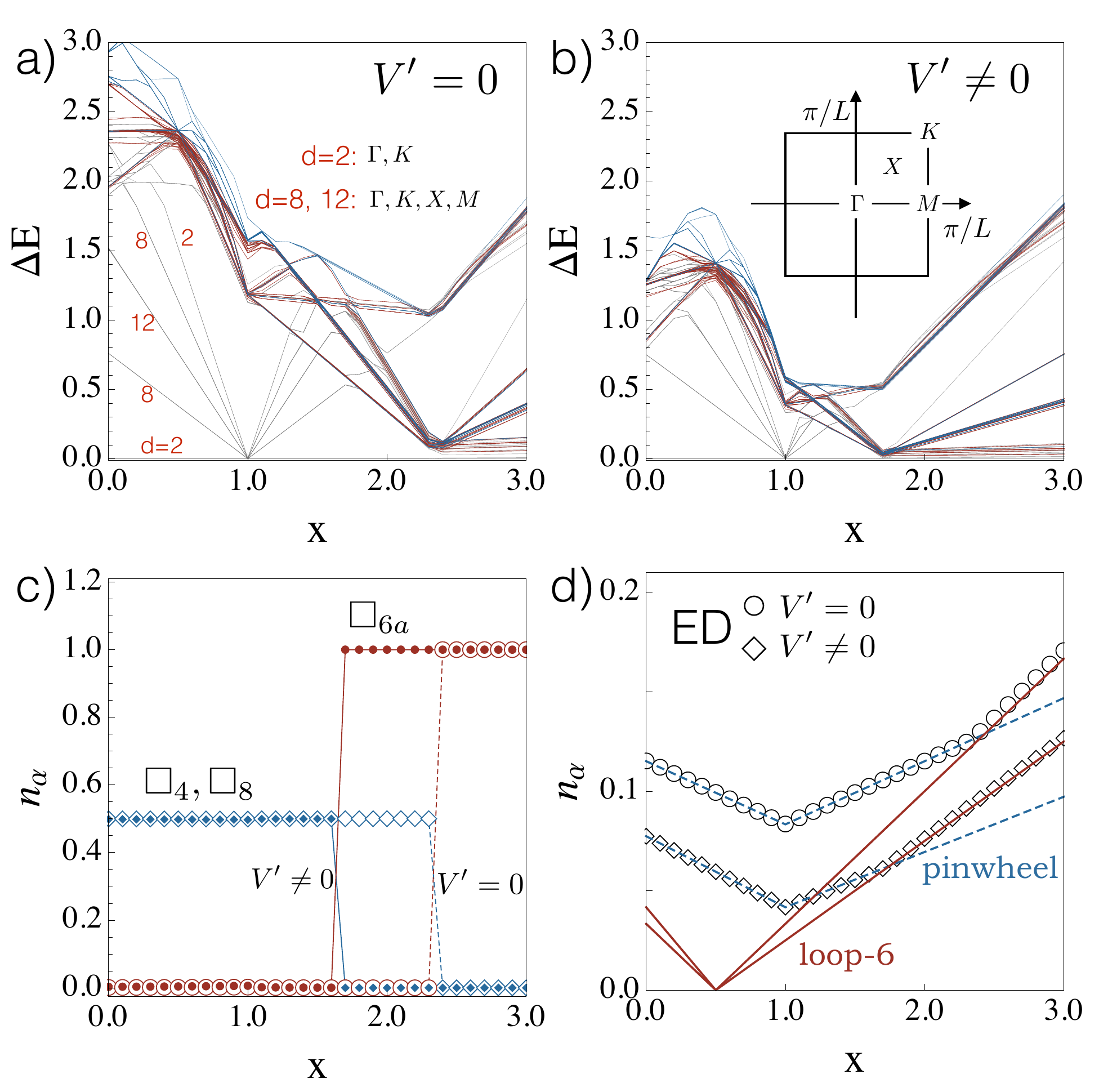}
\caption{(Color online) ED results of (\ref{eq:QDM}) using the minimal NNVB parameters, as a function of $x\!=\!J_{\text{AB}}/J_{\text{AA}}$, with and without potential terms $V'$. (a-b) Low-energy spectra measured from the GS. The 32 states ($2\!\times\!2^{\frac{N}{12}}$) that become degenerate at $x\!=\!1$ (see text) are shown in grey, and the number $d$ is the degeneracy of each level. (c) GS expectation values of various loop densities. (d) Comparison between the GS energy with (\ref{eq:PWL6a}), see text.}\label{fig:3}
\end{figure}

Although the RVB description is only relevant around $x\!=\!1$, it is useful to examine the phase diagram as a function of $x$, in order to establish the main competing phases of the problem. Figure~\ref{fig:3} shows exact diagonalization (ED) results from the 48-site cluster on a torus, using the parameters from the above $2\!\times\!2$ NNVB truncation. The low-energy spectra (a-b) and ground state (GS) expectation values of various loop densities (c) reveal three VBC phases: i) the `even pinwheel' of Fig.~\ref{fig:1}\,(c, left) for $x\!<\!1$, ii) the `odd pinwheel' of Fig.~\ref{fig:1}\,(c, left) for $1\!<\!x\!<\!1.75$, and iii) the `loop-six' VBC of Fig.~\ref{fig:1}\,(c, right) for $x\!>\!1.75$. The potential terms do not alter this picture qualitatively, as shown in Fig.~\ref{fig:3}. Without potential terms the transition between the `odd pinwheel' and the `loop-six' crystals shifts to $x\simeq2.34$.  

All three states belong to the trivial topological sector, and break the translation symmetry in the same way, with a doubling of the unit cell. This symmetry breaking manifests in the spectra by an almost exact degeneracy between two states with momenta $0$ and $(\pi,\pi)$. The extremely small splitting [not visible in Fig.~\ref{fig:3}\,(a,b)] shows that finite-size effects are already very weak for the 48-site cluster. 

Let us look at these states in more detail. The two `pinwheel' states maximize the number of $\Sq_4$ resonances with even parity (i.e. with the symmetric combination of the two NNVB states involved, since $t_{\Sq_4}\!=\!-1\!<\!0$~\footnote{Our singlet-orientation convention is clockwise in void squares and octagons.}), denoted by $|\text{PR}_+\rangle$ in Fig.~\ref{fig:1}\,(c, left). This amounts to a percentage of $\frac{1}{2}$ resonating AA-squares. The remaining half carry `pinwheels', denoted by $|\text{PW}_{\pm}\rangle$, whose parity is fixed by the sign of $t_{\Sq_8}\!=\!\frac{8}{21}(1-x)$, i.e., it is odd (even) for $x\!<\!1$ ($>\!1$). At $x\!=\!1$, the two parities give rise to an extensive degeneracy of $2^{\frac{N}{12}}$ (on top of the two-fold degeneracy from  translational symmetry breaking), which can be clearly seen in the exact spectra of Fig.~\ref{fig:3}\,(a,b). This is analogous to the 36-site VBC of the kagome~\cite{MarstonZeng91,NikolicSenthil03,SinghHuse07,Poilblanc2010,Vidal2010}, which maximizes the `perfect hexagon' resonances, and features a similar extensive degeneracy due to `hexagonal pinwheels'. Here the degeneracy is lifted for $x\!\neq\!1$ due to the inequivalent bond structure~\cite{IoannisKagomeLike}.

Turning to the `loop-six' crystal, this state maximizes the $\Sq_{6a}$ resonances, denoted by $|\text{L}6_-\rangle$ in Fig.~\ref{fig:1}\,(c, right). Here all AA-squares carry such a resonance and the parity is odd since $t_{\Sq_{6a}}\!=\!\frac{1}{5}\!>\!0$. We emphasize here that at the level of the minimal NNVB truncation, $t_{\Sq_{6a}}\!=\!t_{\Sq_{6b}}$, yet the favored `loop-six' phase features only resonances of the type $\Sq_{6a}$. The latter are therefore selected by weak quantum fluctuations driven by the remaining tunneling terms in the minimal QDM. Indeed, states featuring $\Sq_{6b}$ resonances or combinations of $\Sq_{6a}$ and $\Sq_{6b}$ have a very small excitation energy, as can be seen in the low-energy excitation spectrum of Fig.~\ref{fig:3}\,(a,b).  

Fluctuations are in fact very weak for all three VBC phases. As shown in Fig.~\ref{fig:3}\,(c), the GS loop densities of $\Sq_4$ and $\Sq_8$ are almost exactly equal to 1/2 for $x\!<\!1.75$, while that of $\Sq_{6a}$ is almost exactly equal to 1 for $x\!>\!1.75$. The same fact is also demonstrated in Fig.~\ref{fig:3}\,(d) which shows that the exact energies of the lowest eigenstates (symbols) lie practically on top of the corresponding energies (lines)  
\bea\label{eq:PWL6a}
&&
E_{\text{pinwheel}}^{(0)}/N=\left( V_{\Sq_4}-|t_{\Sq_4}| + V_{\Sq_8} -|t_{\Sq_8}| \right)/12~,\nonumber\\
&&
E_{\text{loop-six}}^{(0)}/N=\left( V_{\Sq_{6a}}-|t_{\Sq_{6a}}|\right)/6~,
\eea
of the ideal product states between the resonating building blocks without extra quantum fluctuations. The latter are very weak because they arise from sequences of longer loop events that necessarily involve the octagons, whose tunneling amplitudes are very small, see \cite{IoannisKagomeLike}.

{\it `Dressed' NNVB description} --
Having established the main competing phases of the problem we now focus again on the highly frustrated point $x\!=\!1$.  As announced above, the longer-range singlets that are virtually excited around the defect triangles destabilize the `pinwheel' crystals in favor of the resonating `loop-six' state, i.e. they effectively shift the boundary between the two phases in a drastic way. Essentially, the ratio $-t_{\Sq_4}/t_{\Sq_{6a}}$ drops from the NNVB value of $5$ below $2$, which is the transition value obtained from (\ref{eq:PWL6a}) with $t_{\Sq_8}\!=\!0$ and by neglecting the potential terms (see below). 

Let us see how such a drastic reduction takes place here. Following \cite{IoannisZ2}, we extract the QDM parameters from the ED spectra of appropriate finite Heisenberg clusters, using the lowest two exact eigenvalues with the right symmetry. This method is essentially equivalent to enlarging the truncation basis in order to include the virtual singlets accommodated by the clusters. Figure~\ref{fig:2} shows the clusters considered here for the four processes around AA-squares (for the octagon processes see \cite{SM}). The clusters shown at the top line capture the virtual singlets that are closest to the central loops. 

To capture the contribution from other virtual singlets we proceed as in \cite{IoannisZ2} and add sawtooth chains of length $R$ at the positions of the defect triangles of the minimal $R\!=\!0$ clusters. These chains provide the only escaping paths of the virtual singlets (or the spinons forming these singlets~\cite{Hao10}), provided we do not encounter extra defect triangles nearby. Such extra defect triangles become first relevant for the loop-six processes, and can be studied by examining different variations of the corresponding clusters for each given $R$, that accommodate zero, one or two extra defect triangles, see Fig.~\ref{fig:2}. The corresponding effect on the octagonal processes is disregarded here since these processes are anyway too weak to affect the low-energy physics quantitatively. 

\begin{table}[!t]
\begin{ruledtabular}
\begin{tabular}{@{}l | l | l | l | l | l@{}} 
  & NNVB & R=0& R=1 & R=2 & R=3 \\
\hline
$\Sq_4$  					&  -1 		& -1			& -0.385403	& -0.208442 	& -0.133124	\\
\hline
$\Sq_{6a}^{(0)}$	& +1/5 		& +0.266044		& +0.151461	& +0.125542 	& +0.113368\\		
$\Sq_{6a}^{(1)}$	& +1/5     		& +0.266044         	& +0.131499	& +0.102100	& +0.089369\\
$\Sq_{6a}^{(2)}$	&  +1/5    		&  +0.266044          	& +0.124130	& +0.093471	& +0.081068\\
\
$\Sq_{6b}^{(0)}$	& +1/5		& +0.116951  	& +0.097390	& +0.088453 	& +0.082762\\
$\Sq_{6b}^{(1)}$	& +1/5 		& +0.116951 	& +0.063010	& +0.057934 	& +0.054884\\
$\Sq_{6b}^{(2)}$	& +1/5 		& +0.116951 	& +0.002895	& +0.029735 	& +0.030038\\
$\Sq_8$     & 0 		&  0 			& 0			& 0			& 0\\
\end{tabular}
\end{ruledtabular}
\caption{Tunneling parameters $t$ for the square processes (for the octagonal processes see \cite{SM}), as extracted from the finite spin-1/2 Heisenberg clusters of Fig.~\ref{fig:2}, as described in the text. The numbers from the minimal NNVB truncation are also shown for comparison.}
\label{tab:ts}
\end{table}

The extracted tunneling amplitudes for the square and the octagonal processes are given in Table~\ref{tab:ts} and Table~2 of \cite{SM}, respectively, and deliver a number of clear insights (the potential terms are much smaller and can be disregarded~\cite{SM}). First, except for the minimal $R=0$ embedding of $t_{\Sq_4}$ (and the processes $\Sq_8$ and $\Oct_{16}$ whose amplitudes are zero irrespective of $R$), the amplitudes are clearly different from the ones resulting from the minimal $2\!\times\!2$ truncation (NNVB). The differences show that  the $2\!\times\!2$ truncation give in most cases a very poor approximation, so the virtual singlets cannot be ignored. 

Second, the shortest amplitude $t_{\Sq_4}$ shows a remarkable reduction by about {\it ten} times from $R\!=\!0$ to $R\!=\!3$. By contrast, the amplitudes $t_{\Sq_{6a}}^{(e)}$ drop by about 2-3 times (depending on $e$), and show a relatively faster convergence with $R$ compared to $t_{\Sq_4}$. Virtual singlets have therefore a much stronger tendency to penalize the shortest loop events because they correspond to the highest local density of defect triangles. With the ratio $-t_{\Sq_{4}}/t_{\Sq_{6a}}^{(e)}$ dropping below the transition value of $2$ above $R\!=\!2$, it is immediately clear that the RVB physics will no longer be governed by the shortest resonances.

\begin{figure}[!t]
\includegraphics[width=0.35\textwidth,clip]{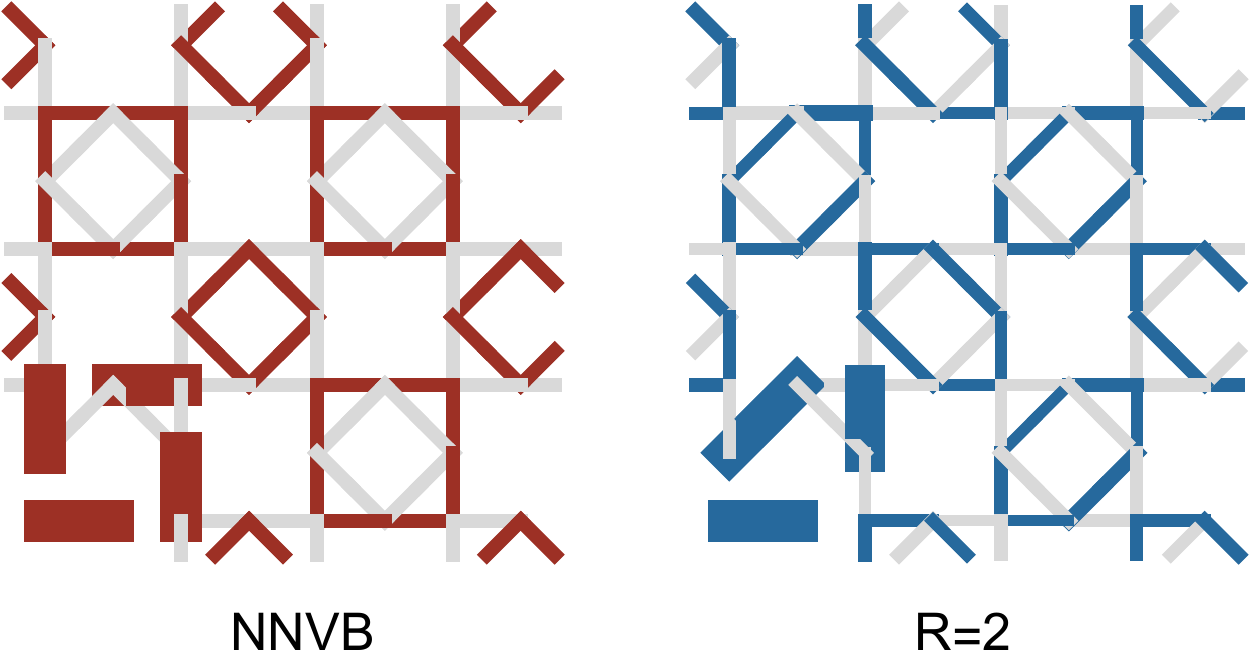}
\caption{(Color online) Connected dimer-dimer correlation profiles in the GS of (\ref{eq:QDM}) at $x\!=\!1$, using the NNVB tunneling parameters (left) and the ones extracted from the $R\!=\!2$ clusters of Fig.~\ref{fig:2} (right), which includes the explicit dependence on the extra defect triangles (see text). The reference dimer is the horizontal segment at the bottom left corner. The red and grey (left) or blue and grey (right) denote positive and negative  values.}\label{fig:4}
\end{figure}

Third, our results show a $\sim\!20\%$ variation of $\Sq_{6a}^{(e)}$ and a much stronger $\sim\!40$-$60\%$ variation of $\Sq_{6b}^{(e)}$ depending on $e$, i.e. on whether there are zero, one or two extra defect triangles nearby. Such a strong dependence on one class of the dominant processes of the problem must be taken into account explicitly in the effective description. The presence of such correlated loop processes, which depend strongly not only on the type of the loop but also on the embedding, highlights the qualitatively non-trivial impact of virtual singlets, which effectively sample the immediate VB environment of a loop.

Fourth, virtual singlets have a different impact on $t_{\Sq_{6a}}$ and $t_{\Sq_{6b}}$, again in contrast to the minimal NNVB truncation. For all $R$ we systematically find that $t_{\Sq_{6a}}^{(e)}\!>\!t_{\Sq_{6b}}^{(e)}$, showing a clear preference for loop-six resonances of the $\Sq_{6a}$ type. In the minimal NNVB model discussed above, such events were selected by weak virtual octagonal processes whereas here they are favored directly by virtual singlets. 

Finally, the processes around octagons are also renormalized substantially by virtual singlets, see \cite{SM}. But, these processes remain anyway harmless, which is checked by a full ED of the `dressed' QDM including all 22 processes.

Fig.~\ref{fig:4} shows the connected dimer-dimer (or density-density in the QDM setting) GS correlation profiles using the parameters from the minimal NNVB truncation and from the $R=2$ clusters, with the explicit embedding dependence for the loop-six processes. At this level, we have access to all amplitudes except $t_{\Oct_8}$, whose corresponding cluster is too large for ED. In the calculations shown here we have taken the conservative assumption $t_{\Oct_8}^{R=2}=t_{\Oct_8}^{R=1}$, which does not affect the results. The data confirm that the `loop-six' crystal is stabilized already at $R\!=\!2$, and taking the $R\!=\!3$ amplitudes renders this phase even more robust.

{\it Outlook} --
We have shown that the minimal truncation to the NNVB basis fails qualitatively even in a system where it is expected to work the most, according to received wisdom. This is strong evidence that  virtual singlets have a qualitative impact in all standard disordered AFMs. Not only do the virtual singlets `dress' the basic RVB building blocks, they also completely alter fundamental notions that derive from our experience with the minimal NNVB truncation: The tunneling amplitudes {\it do not} drop exponentially in the length $L$ of the loops, and the tunneling dynamics cannot always be approximated as a sum over strictly local loop processes, as initially envisaged by Rokhsar and Kivelson~\cite{RokhsarKivelson}.

{\it Acknowledgements} -- We are grateful to F. Mila for enlightening discussions and suggestions. This work is supported by the French National Research Agency through Grants No. ANR-2010-BLANC-0406-0 NQPTP.


%

\clearpage

\appendix

\pagenumbering{roman}
\widetext

\begin{widetext}
\section{Supplemental material}

\subsection{A. General remarks for the NNVB manifold of the square-kagome antiferromagnet}
As shown in [\onlinecite{IoannisKagomeLikeS}] by a generalization of the so-called arrow representation~\cite{Elser1989S}, the dimension of the NNVB Hilbert space is $2\!\times\!2^{N/3}$, where $N$ is the total number of sites. Similar to the non-bipartite, triangular and kagome lattices, the NNVB manifold splits into four topological sectors for periodic boundary conditions in both directions. These sectors can be labeled by the parities ($p_x$,$p_y$) of the number of dimers intersecting any horizontal or vertical line~\cite{Moessner2001S,Fendley2002S,Ioselevich2002S,Ralko2005S}. Here, the sectors (0,1) and (1,0) are connected to each other by the four-fold symmetry of the lattice.


\begin{table*}[b]
\ra{1.3} 
\begin{ruledtabular}
\begin{tabular}{@{}r|lllllll@{}} 
void plaquette & process & loop length $L_p$ & $\omega_p$ & ${\sf v}_p/\omega_p$ & $E_{0,p}$ & $t_p$ & $V_p$ \\
\hline
AA-square &$\Sq_{4}$& $4$ & $+2^{-1}$ & $-3$ & $-\frac{3}{2}$ & $-1$ & $+\frac{1}{2}$\\
                  &$\Sq_{6a,b}$ & $6$ & $-2^{-2}$ & $-3x$ & $-\frac{3}{4}(2x+1)$ & $+\frac{1}{5}(2x-1)$ & $+\frac{1}{20}(2x-1)$\\
                  & $\Sq_{8}$& $8$ & $+2^{-3}$ & $-3(2x-1)$ & $-3x$ & $+\frac{8}{21}(1-x)$ & $-\frac{1}{21}(1-x)$\\
\hline
AB-octagon &$\Oct_{8}$& $8$   & $+2^{-3}$ & $-6x$ & $-3x$ & $-\frac{8}{21}x$ & $+\frac{1}{21}x$\\                    
                   & $\Oct_{10a-d}$& $10$ & $-2^{-4}$ & $-\frac{3}{2}(3x+1)$ & $-\frac{3}{4}(4x+1)$ & $+\frac{4}{85}(2x+1)$ & $+\frac{1}{340}(2x+1)$\\       
                   &$\Oct_{12a-f}$ & $12$ & $+2^{-5}$ & $-3(x+1)$ & $-\frac{3}{2}(2x+1)$ & $-\frac{16}{341}$ & $+\frac{1}{682}$\\ 
                   & $\Oct_{14a-d}$& $14$ & $-2^{-6}$ & $-\frac{3}{2}(x+3)$ & $-\frac{3}{4}(4x+3)$ & $+\frac{16}{1365}(3-2x)$ & $+\frac{1}{5460}(3-2x)$\\ 
                   & $\Oct_{16}$& $16$ & $+2^{-7}$ & $-6$ & $-3(1+x)$ & $-\frac{128}{5461}(1-x)$ & $+\frac{1}{5461}(1-x)$
\end{tabular}
\end{ruledtabular}
\caption{QDM parameters $t_p$ and $V_p$ (in units of $J_{\sf AA}$) for the four processes around AA-squares and the 18 processes around AB-octagons, using the minimal $2\!\times\!2$ NNVB truncation approach. $L_p$ is the length of the loop in the transition graph of $|1_p\rangle$ and $|2_p\rangle$. For the dimer orientations we follow the convention that singlets are oriented clockwise in each even-length loop. All values correspond to the clusters of the first row of Fig.~\ref{fig:supp1}. If we use clusters of any other row, $E_{0,p}$ and ${\sf v}_p$ will change, but the overlap $\omega_p$ and the combination $h_p=v_p/\omega_p-E_{0,p}$ do not change, leading to the same $t_p$ and $V_p$.}
\label{tab:qdm}
\end{table*}

\subsection{B. Basic elements of the minimal $2\!\times\!2$ NNVB truncation}
The basic elements of the minimal $2\!\times\!2$ NNVB truncation method in the square-kagome AFM are given in [\onlinecite{IoannisKagomeLikeS}]. For convenience they are repeated here in Table~\ref{tab:qdm}.  
Let us consider any of the finite spin-1/2 Heisenberg clusters of the first row of Fig.~(2) of the main text (or Fig.~\ref{fig:supp1} for the octagonal processes) and consider the elementary tunneling process $p$ between the two dimer coverings $|1_p\rangle$ and $|2_p\rangle$ that can be accommodated by this cluster. For the dimer orientations~\cite{SMfootnote}, we take the singlets to be oriented clockwise in each void square and octagon. The transition graph~\cite{Sutherland88S} of our elementary process involves a single non-trivial loop of length $L_p$, surrounding a single square or a single octagon. Then, within our convention, the overlap between the two NNVB states is given by
\be\label{eq:omega1}
\omega_p =  \langle 1_p|2_p\rangle = (-1)^{\frac{L_p}{2}}~2^{1-\frac{L_p}{2}} ~.
\ee
The diagonal and off-diagonal matrix elements of the Hamiltonian, 
\be
E_{0,p}=\langle 1_p | \mc{H}_{\text{Heis}} | 1_p\rangle=\langle 2_p | \mc{H}_{\text{Heis}} | 2_p\rangle, ~~~v_p=\langle 1_p | \mc{H}_{\text{Heis}} | 2_p\rangle,
\ee
can be found using standard rules~\cite{Sutherland88S, RokhsarKivelsonS, ZengElser95S, MambriniMila2000S, Misguich02S, Misguich03S, Ralko2009S}.
In the basis $\{|1_p\rangle, |2_p\rangle\}$ we have
\be
\mc{H}_{\text{NNVB}}=(\mc{O}^{-\frac{1}{2}}\mc{H}_{\text{Heis}}\mc{O}^{-\frac{1}{2}})_{\text{NNVB}}=
\left(\begin{array}{cc}
1 & \omega_p\\
\omega_p & 1
\end{array}\right)^{-\frac{1}{2}}
\left(\begin{array}{cc}
E_{0,p} & v_p\\
v_p & E_{0,p}
\end{array}\right)
\left(\begin{array}{cc}
1 & \omega_p\\
\omega_p & 1
\end{array}\right)^{-\frac{1}{2}}
=\left(\begin{array}{cc}
E_{0,p}+V_p & t_p\\
t_p & E_{0,p}+V_p
\end{array}\right)
\ee 
with
\be\label{eq:tV}
t_p = + h_p \frac{\omega_p}{1-\omega_p^2},~~V_p=-t_p ~\omega_p,~~\text{where}~h_p=v_p/\omega_p-E_{0,p}~.
\ee
The quantity $h_p$ corresponds to Eq.~(12) of Schwandt {\it et al}~\cite{Schwandt2010S}, modulo a factor of 3/4 related to the redefinition of the Hamiltonian; see passage below Eq.~(10) in [\onlinecite{Schwandt2010S}]. So Eq.~(\ref{eq:tV}) is consistent with Eq.~(40) of [\onlinecite{Schwandt2010S}].  
In other words, the above $2\!\times\!2$ truncation is equivalent with the infinite-order cluster expansion of [\onlinecite{Schwandt2010S}].

\subsection{C. Potential terms}
Here we discuss the potential terms of the effective QDM. As explained in [\onlinecite{IoannisZ2S}], the largest contributions to the potential energy arise from processes involving a single defect triangle, and as such they can be absorbed in a global energy shift, since the total number of defect triangles is a constant in the NNVB basis. The remaining contributions can be divided into binding energies among two, three, etc. defect triangles. In the present case, the most important processes (i.e. the ones around the AA-squares) involve up to two nearby defect triangles, for which the binding energy is about 0.01~\cite{IoannisZ2S}, i.e. 10 times smaller than the dominant tunneling amplitudes. This is why we can safely disregard the potential terms in the effective QDM description of the square-kagome.

\subsection{D. Finite clusters for the octagonal processes}
Figure~\ref{fig:supp1} shows the finite-size spin-$\frac{1}{2}$ Heisenberg clusters that are used to extract the tunneling parameters of the 18 different octagonal processes (for the processes around AA-squares see main text).

\begin{figure*}[!h]
\includegraphics[width=0.52\textwidth,clip]{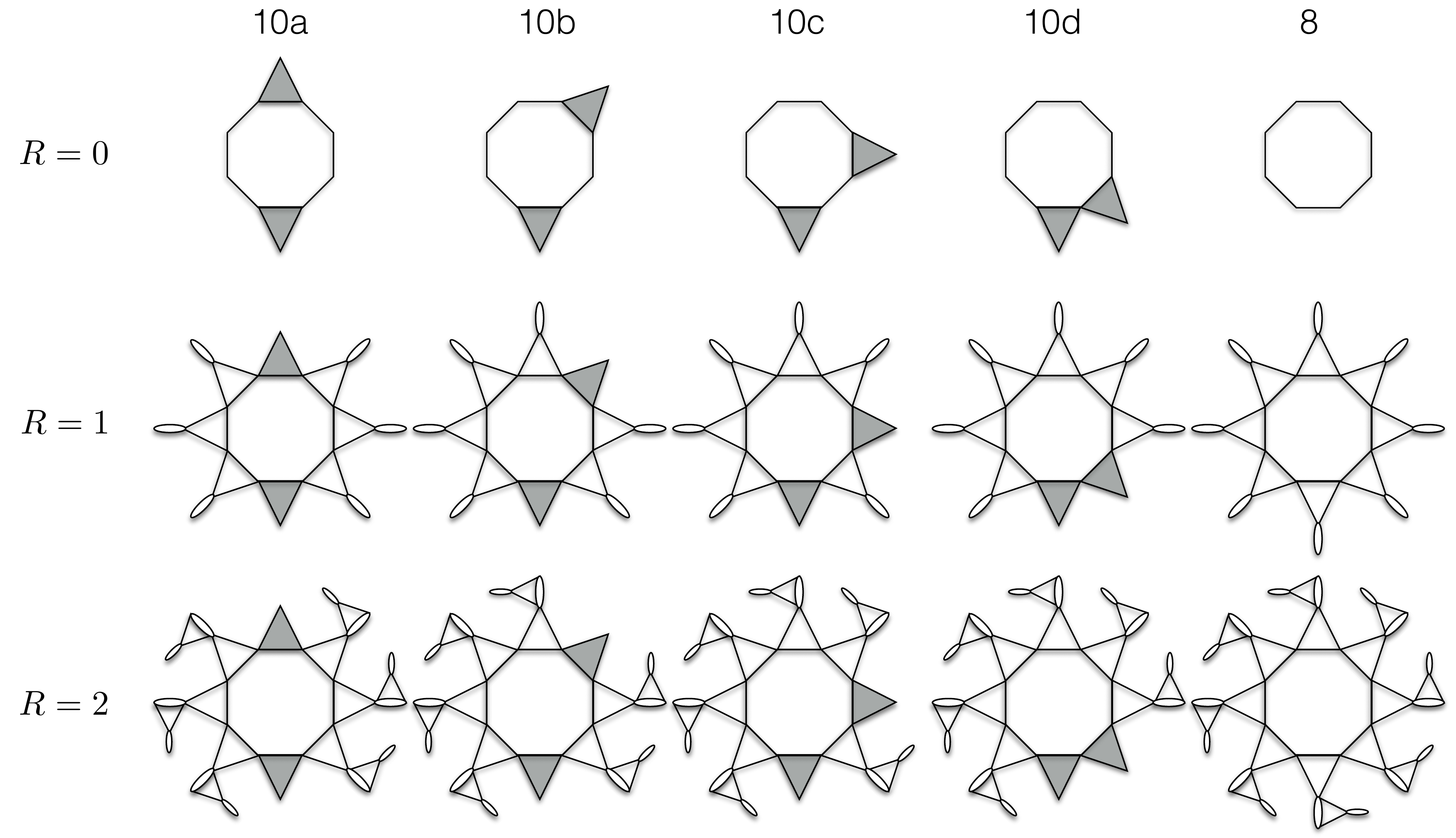} \\
\vspace{8px}
\includegraphics[width=0.52\textwidth,clip]{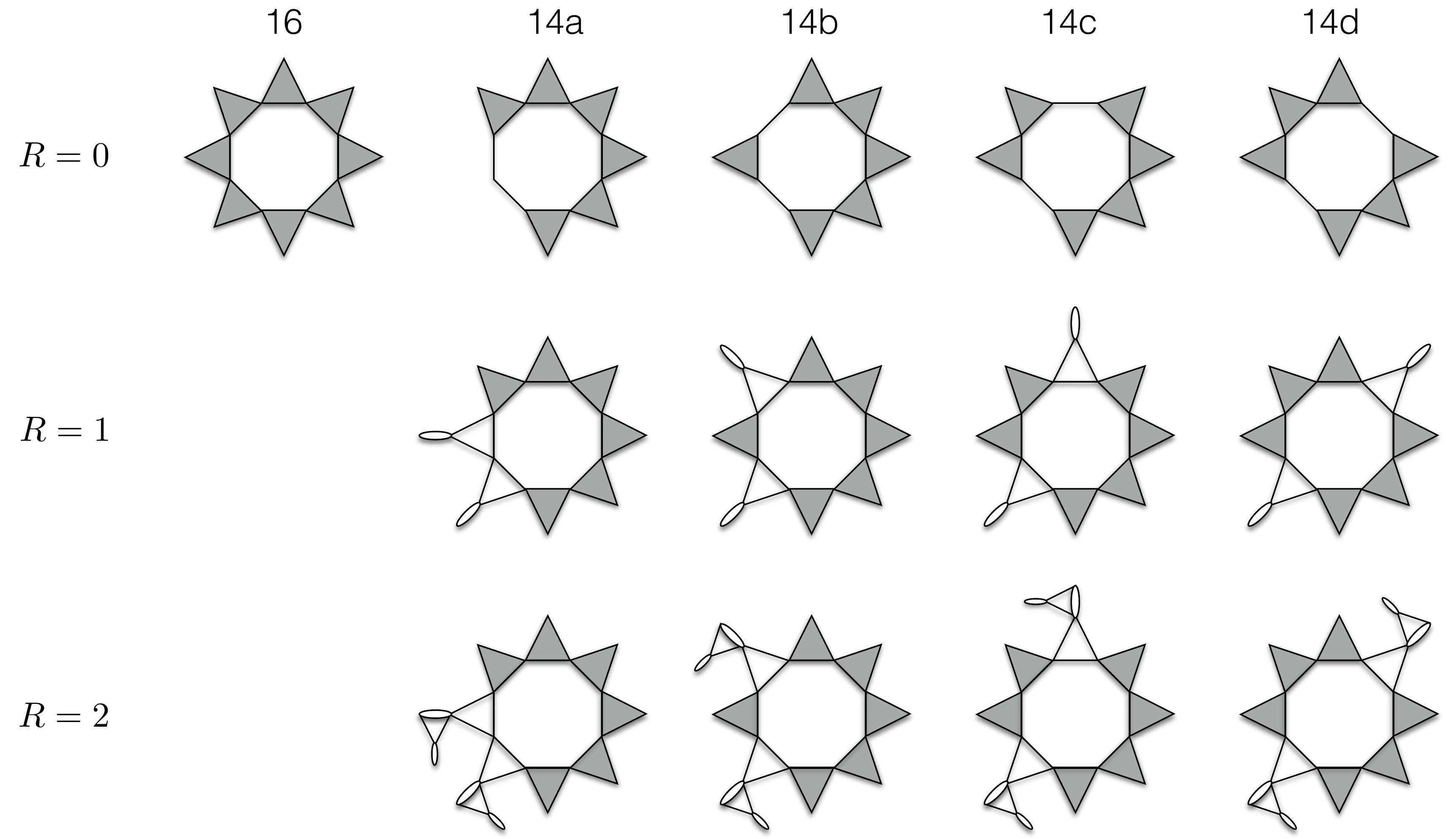} \\
\vspace{8px}
\includegraphics[width=0.50\textwidth,clip]{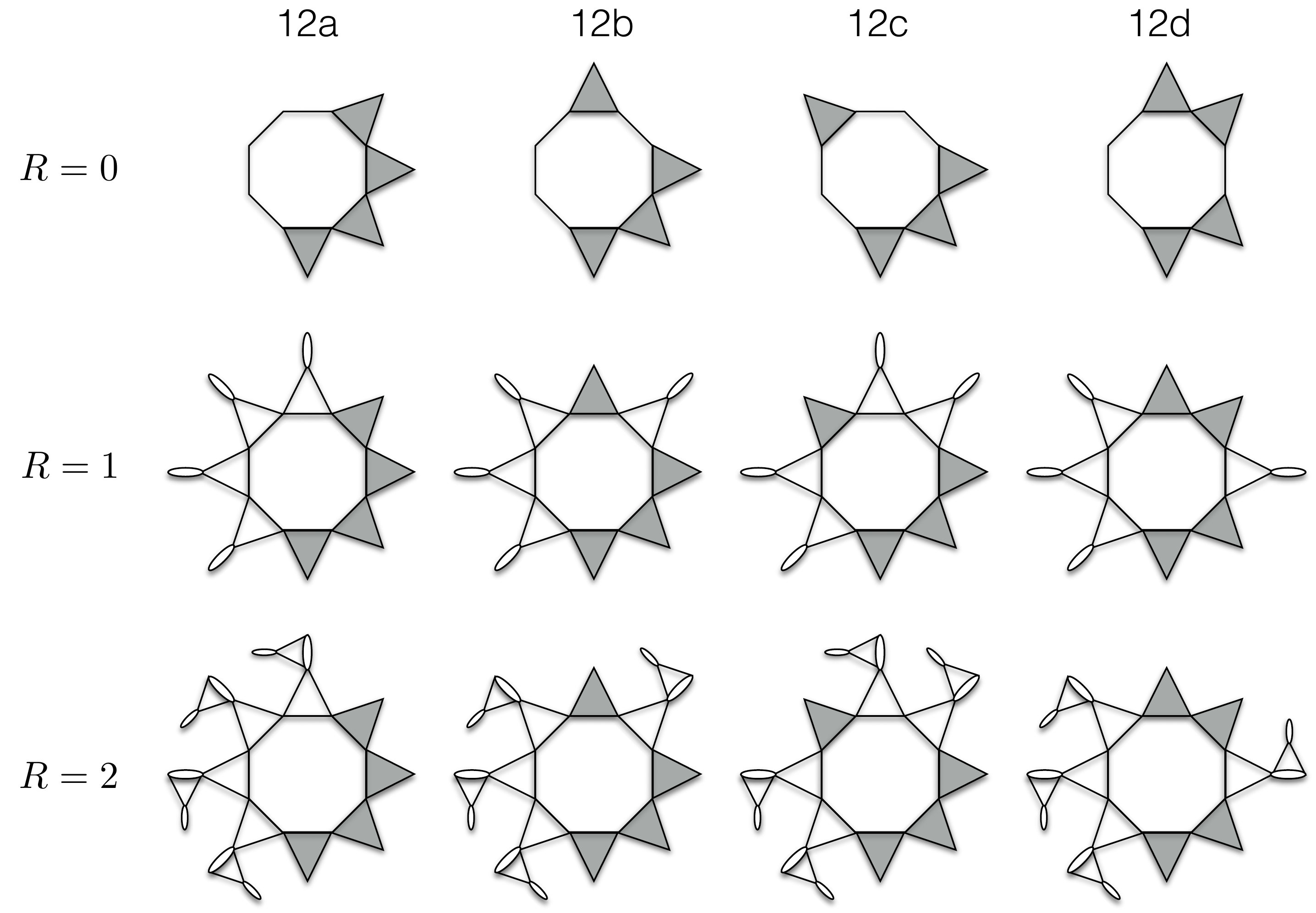} \\
\vspace{8px}
\includegraphics[width=0.50\textwidth,clip]{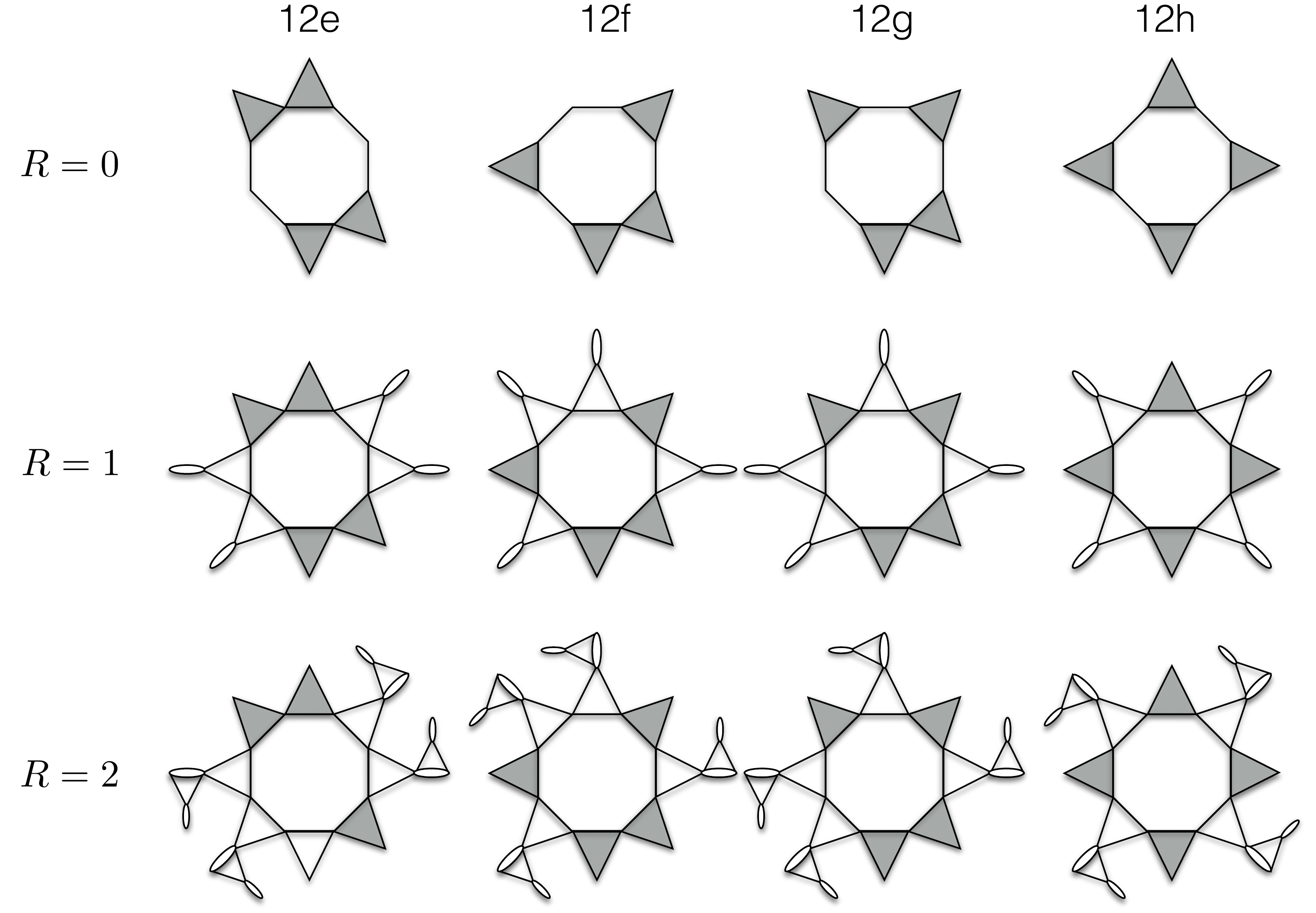}
\caption{Finite clusters used to extract the octagonal tunneling parameters of the effective Hamiltonian. The gray triangles are the defect triangles involved in each tunneling process. 
}\label{fig:supp1}
\end{figure*}


%

\end{widetext}

\end{document}